\documentclass[final,5p,times,twocolumn]{elsarticle}

\usepackage{amssymb}

\usepackage{newfloat}
\DeclareFloatingEnvironment[name={Supplementary Figure},fileext=lsf,listname={List of Supplementary Figures}]{suppfigure}

\usepackage[backref]{hyperref}

\journal{The Ocular Surface}

\begin{document}

\begin{frontmatter}



\title{AI-based automated Meibomian gland segmentation, classification and reflection correction in infrared Meibography}


\author[Gist_biomed]{Ripon Kumar Saha}
\author[Gist_biomed]{A. M. Mahmud Chowdhury}
\author[Yeouido]{Kyung-Sun Na}
\author[Yeouido]{Gyu Deok Hwang}
\author[Korea_U]{Youngsub Eom}
\author[Chungnam]{Jaeyoung Kim}
\author[GISTAI]{Hae-Gon Jeon}
\author[Yeouido]{Ho Sik Hwang*}
\author[Gist_biomed,GISTAI]{Euiheon Chung*}

\affiliation[Gist_biomed]{organization={Department of Biomedical Science and Engineering, Gwangju Institute of Science and Technology},
            addressline={Gwangju},
            country={Korea}}
            
\affiliation[Yeouido]{organization={Department of Ophthalmology, Yeouido St. Mary's Hospital, College of Medicine, The Catholic University of Korea},
            addressline={Seoul},
            country={Korea}}
            
\affiliation[Korea_U]{organization={Department of Ophthalmology,Korea University College of Medicine},
            addressline={Seoul},
            country={Korea}}

\affiliation[Chungnam]{organization={Department of Ophthalmology, Chungnam National University School of Medicine},
            addressline={Daejeon},
            country={Korea}}
            
\affiliation[GISTAI]{organization={AI graduate school, Gwangju Institute of Science and Technology},
            addressline={Gwangju},
            country={Korea}}


\begin{abstract}
\textit{\textbf{Purpose:}} Develop a deep learning-based automated method to segment meibomian glands (MG) and eyelids, quantitatively analyze the MG area and MG ratio, estimate the meiboscore, and remove specular reflections from infrared images.\\
\textit{\textbf{Methods:}} A total of 1600 meibography images were captured in a clinical setting. 1000 images were precisely annotated with multiple revisions by investigators and graded 6 times by meibomian gland dysfunction (MGD) experts. Two deep learning (DL) models were trained separately to segment areas of the MG and eyelid. Those segmentation were used to estimate MG ratio and meiboscores using a classification-based DL model. A generative adversarial network was implemented to remove specular reflections from original images.\\
\textit{\textbf{Results:}} The mean ratio of MG calculated by investigator annotation and DL segmentation was consistent 26.23\% vs 25.12\% in the upper eyelids and 32.34\% vs. 32.29\% in the lower eyelids, respectively. Our DL model achieved 73.01\% accuracy for meiboscore classification on validation set and 59.17\% accuracy when tested on images from independent center, compared to 53.44\% validation accuracy by MGD experts. The DL-based approach successfully removes reflection from the original MG images without affecting meiboscore grading.\\
\textit{\textbf{Conclusions:}} DL with infrared meibography provides a fully automated, fast quantitative evaluation of MG morphology (MG Segmentation, MG area, MG ratio, and meiboscore) which are sufficiently accurate for diagnosing dry eye disease. Also, the DL removes specular reflection from images to be used by ophthalmologists for distraction-free assessment. 
\end{abstract}




\end{frontmatter}


\section{Introduction}

Meibomian gland dysfunction (MGD) is a chronic, diffuse abnormality of the meibomian glands (MG), commonly characterized by terminal duct obstruction and changes in glandular secretion.\cite{RN1} A diagnosis of MGD requires evaluation of subjective symptoms, morphologic changes in the eyelid margin, meibum secretion, infrared meibography, lipid layer thickness measurement, and other factors.\cite{RN2} Although examining MG secretion and identifying abnormal eyelid margins are the main methods for clinical evaluation of MGD, meibography has become widely used to assess the morphologic changes of the MG since the introduction of transillumination meibography\cite{RN3} and non-contact infrared meibography.\cite{RN4, RN5} Non-contact infrared meibography uses infrared light and an infrared camera to provide MG images with reduced discomfort for the patient. After obtaining images, the MG morphology can be assessed, including MG dropout (MG loss) area, width, length, and tortuosity \cite{RN6}. Among these parameters, the MG dropout area is one of the most critical.\cite{RN4, RN7, RN8, RN9, RN10, RN11} \par

For semi-quantitative analysis, different methods for grading MG dropout have been proposed. Pflugfelder et al.\cite{RN12} used a grading system (Grade 0: no gland dropout, Grade 1: \textless{}33\% gland dropout; Grade 2: 33\%\textendash{}66\% gland dropout; Grade 3: \textgreater{}66\% gland dropout). Nichols et al.\cite{RN13} used meibography image grading scales (Grade 1: no partial gland, Grade 2: \textless{}25\% partial glands; Grade 3: 25\%\textendash{}75\% partial glands; Grade 4: \textgreater{}75\% partial glands). Arita et al.\cite{RN4} defined a meiboscore based on 4 grades, Grade 0: no dropout, Grade 1: dropout of less than 1/3, Grade 2: dropout of more than 1/3 and less than 2/3, and Grade 3: dropout of more than 2/3. Pult et al.\cite{RN13} used degree 0 = no partial glands; 1 = \textless{}25\% partial glands; 3 = 25\%-50\% partial glands; 3 = 50\%-75\%, and 4: \textgreater{}75\% partial glands. \par

For quantitative analysis, some studies measured the MG dropout ratio using image software such as ImageJ.\cite{RN14, RN15, RN16, RN17} In these studies, however, only the contour of normal MGs except the dropout area was defined and MG dropout ratio was measured. They didn't measure the area of the individual MGs. Arita et al.\cite{RN9} objectively evaluated the MG area using custom-developed software with noninvasive meibography. Lid borders were determined, and the MG area was automatically discriminated by enhancing the contrast and reducing image noise. They succeeded in calculating the ratio of the total MG area relative to the total analysis area in the upper and lower eyelids with their software.\cite{RN9} As described in their discussion, however, manual correction was necessary after automatic detection of MGs was applied in images with too much reflected light and excessive MG loss. \par

Traditional lower-level computer vision approaches or classical techniques were recently used to segment MG, such as adaptive threshold with filtering combining morphologic, local entropy, fourier transform, and texture filtering.\cite{RN18, RN19} They often do not work on various ocular images because of the complex gland and eyelid architecture with different contrast and sharpness levels. Another study proposed a novel method to segment eyelids with a curve fitting technique for MG irregularity quantification and estimation of the dropout area based on the MG morphology.\cite{RN20} However, manual selection of the region of interest (ROI) was necessary for some images. \par

Recently, a deep learning (DL)-based method was proposed to segment the overall MG atrophy area, but not the individual MG morphology level.\cite{RN21} Another approach was proposed to divide the MG images into small sections and used a convolutional neural network to segment individual MGs.\cite{RN22} While the authors provided the MG morphology using the simplified convolutional neural network and correlation with clinical parameters, only 60 images (40 training and 20 validation) of right eyes were used, which can result in a skewed distribution with an overfitting problem. Another deep learning model was introduced by Setu et al.\cite{RN23} to find individual gland properties (gland number, gland length, gland width, and tortuosity) by segmenting MG images for objective assessment of morphometric parameters. This approach is able to objectively analyze individual glands in most cases where multiple glands are not connected together and individual gland doesn't appear disconnected in segmentation.\par

Overall, classical methods are insufficient to segment MG or eyelid on different illumination and require manual adjustment for each image. Besides, current deep learning approaches lack important parameters like MG area, MG ratio, meiboscore without extensive analysis on a large dataset. Herein, we present a novel DL approach to segment MG, eyelid, find important parameters like MG area, MG ratio and Meiboscore for fully automated MG assessment. We have represented extensive analysis of our findings based on a dataset of 1,000 meibography images (publicly available: https://mgd1k.github.io/) with precise segmentation and 6 sets of meiboscore in correlation with age, sex, MG area/ratio and meiboscore. Furthermore, we validated our original model to another dataset of 600 images from a different test center by several methods, including comparisons between meiboscores obtained by MGD experts and DL segmentation with MG area/ratio.\par

\section{Patients and methods}
\subsection{Patients}
This retrospective study was approved by the Institutional Review Board of Yeouido St. Mary's Hospital (SC20RISE0063) and Korea University Ansan Hospital (2022AS0015) and adhered to the tenets of the Declaration of Helsinki. The study included 572 eyes of 320 patients who visited Yeouido St. Mary's Hospital for a dry eye examination. The inclusion criteria were age $\ge20$ years and at least mild dry eye symptoms\cite{RN24} (Ocular Surface Disease Index [OSDI] score $>=13$) and low tear film break-up time (TBUT $<5$ s) using fluorescein eye, a low Schirmer I score ($<10$ mm per 5 min without anesthesia), or corneal punctate fluorescein staining (Oxford staining score of $>1$) in at least 1 eye.\cite{RN25} Exclusion criteria were (1) history of ocular injury; (2) eyelid infection; (3) ocular surgery within the previous 6 months; (4) non–dry-eye ocular inflammation; (5) uncontrolled systemic disease; and (6) unable to undergo infrared meibography. \par

We created a dataset of 1000 MG infrared images named `MGD-1K'. The MGD-1K dataset comprised 1000 images from 320 patients (467 images from the upper eyelids and 533 images from the lower eyelids). The infrared MG images were obtained with the LipiView interferometer (LipiView II Ocular Surface Interferometer, Johnson \& Johnson Inc., Jacksonville, FL). We used only non-contact infrared images, and no transilluminated images. The examination was conducted by one examiner. The dataset is open-source and can be accessed online (https://mgd1k.github.io). \par

\subsection{Segmenting images of MG and eyelid by investigators: Annotation}

\begin{figure}
    \centering
    \includegraphics[width=.70\columnwidth]{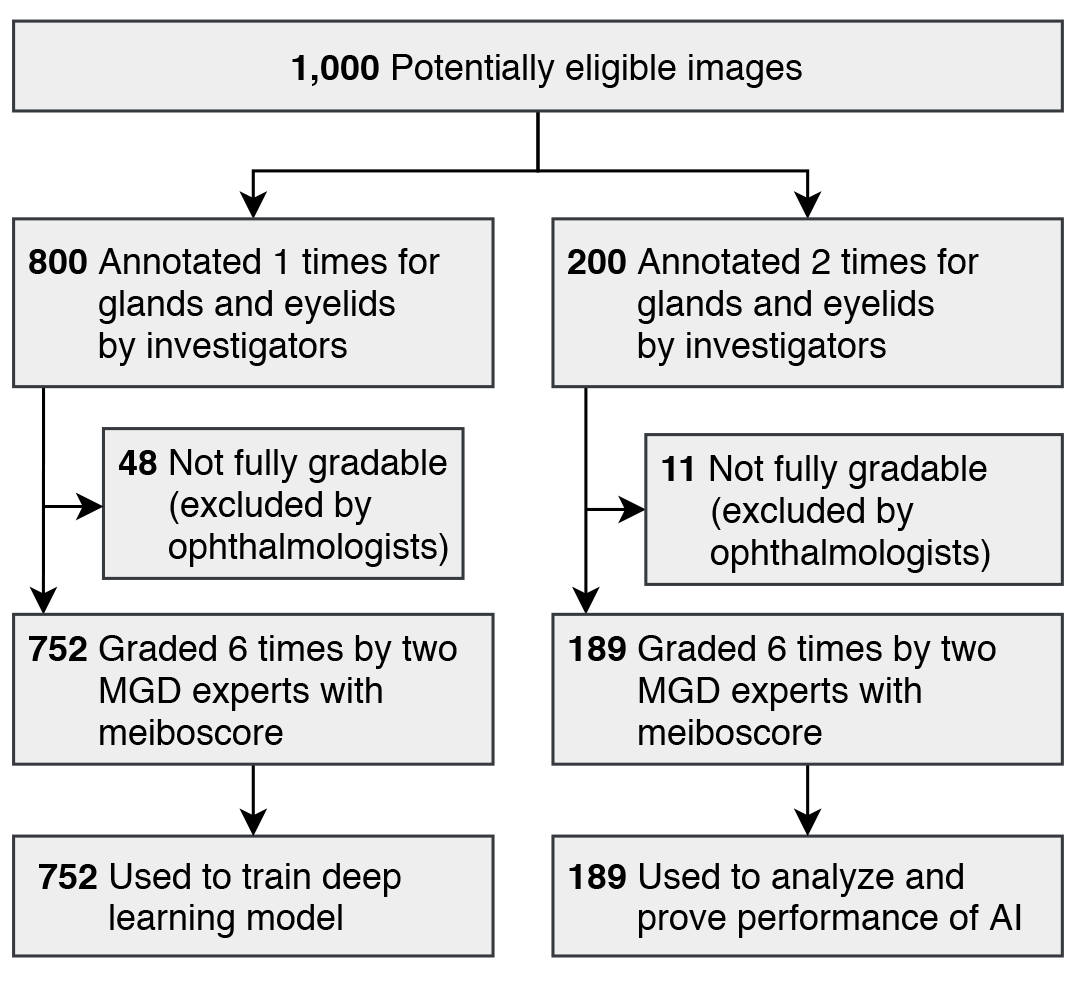}
    \caption{Flowchart showing how a total of 1000 infrared images contained in our dataset MGD-1K were processed. All images were annotated with the location of the meibomian gland and eyelids. In addition, all images were graded in 6 rounds by 2 meibomian gland dysfunction experts (ophthalmologists). A total of 59 images were marked as ungradable by at least by 1 ophthalmologist in at least 1 round. The remaining 941 images were divided into training images and validation images to train the deep learning model and to analyze the performance.}
    \label{Figure-1.png}
\end{figure}

\begin{figure}
    \centering
    \includegraphics[width=.99\columnwidth]{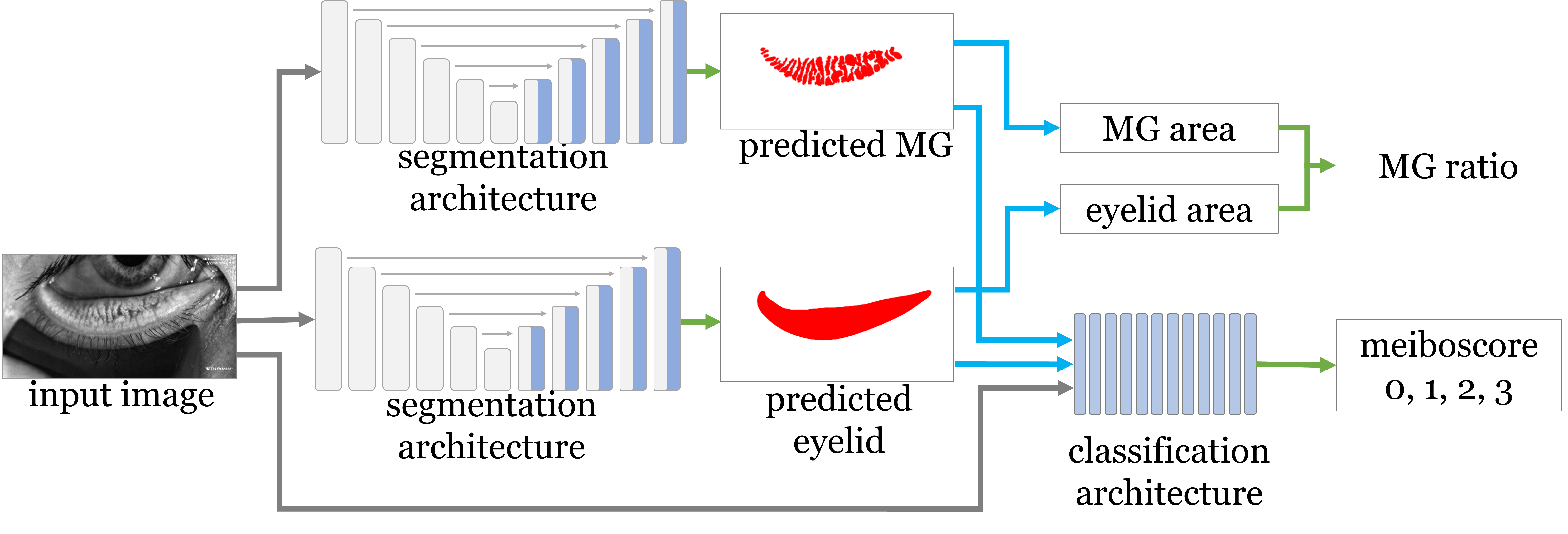}
    \caption{Representation of the overall deep learning model comprising 3 models with segmentation blocks added to provide meibomian gland (MG) segmentation and eyelid segmentation for each, and a classification block to predict the meiboscore. Initially, 2 separate segmentation blocks were trained with original images corresponding to their segmentation. The segmentation model was based on an encoder-decoder architecture with skip connections between them where the encoder is based on Resnet34.\cite{RN33} The initial 2 segmentation blocks produce a meibomian gland segmentation image and an eyelid segmentation image as output from the original image. The MG area and eyelid area can be calculated from these segmentation. The MG ratio can also be calculated by dividing the MG area by the eyelid area. The last block works as a classifier, which takes the original input image and previous layers of gland and eyelid segmentation as input to predict the meiboscore.}
    \label{Figure-2.png}
\end{figure}

All 1000 infrared images were labeled by drawing over the MGs and eyelid by a team of investigators (computer scientists) in close observation and direct supervision by an MGD expert (H.S.H.) (Fig. \ref{Figure-1.png}). Multiple revisions were performed to make those segmentation as precise as possible. The Adobe Photoshop (CC 2019) brush and eraser tool were used to precisely segment each MG. Image pre-processing was performed during segmentation to enhance the visibility of the MGs by adjusting the brightness, contrast, exposure, color inversion, and gamma correction. Later, those segmentations were post-processed to create images consisting of MG/eyelid information to feed into the DL model (Fig. \ref{Figure-2.png}). \par

The eyelid area was defined vertically by the line of the lid margin to the edge of the tarsal fold. This was somewhat subjective for the lower lid because the superior edge of the fold is not defined by a discrete and clearly demarcated tarsal plate.\cite{RN26} The eyelid area was defined horizontally by the lacrimal punctum to the lateral canthus. \par

The MG and eyelid area in the upper and lower eyelid were calculated from the processed segmentations. The MG ratio in the upper and lower eyelids was defined by the ratio of the MG area and the corresponding eyelid area. The mean and standard deviation [SD] of the MG area and ratio in the upper and lower eyelids were obtained. To validate the annotation results obtained by the investigators, we calculated the correlation between the MG area/ratio and age, because it is well known that MG dropout increases with age.\cite{RN4} \par

\subsection{Meiboscore grading by 2 MGD experts}
To train/validate the DL model, 2 MGD experts graded meiboscores of 1000 images of the `MGD-1K' according to Arita's criteria.\cite{RN4} We used Labelbox\cite{RN27} for the meiboscore collection. Before sending the images to the MGD experts, all image names and personal identifiers were removed. All 1000 images were graded 3 times for meiboscore by 2 MGD experts separately; thus, 6 times per image (Fig. \ref{Figure-1.png}). All images were randomly ordered in each round. There was at least a 1-week interval between each round so that the previous grading would not influence the next round of grading. Also, the 2 MGD experts (H.S.H., K.S.N.) who served as graders were not allowed to share their results with each other. Besides providing meiboscores, we included an option for the graders to select any image as ungradable. A total of 59 images were indicated as ungradable in at least 1 round of grading. Thus, the remaining 941 images with corresponding meiboscore were used to train/validate DL. The mean meiboscore was calculated by averaging meiboscore from all graders. However, we got 139 cases (out of 1,000) with average meiboscore 0.5, 45 cases with meiboscore 1.5, and 8 cases with average meiboscore 2.5, and we have rounded up those mean to the nearest integer meiboscore 1,2 and 3, respectively. \par

We examined the variation in the 3 gradings by each MGD expert. Also, to determine the grading consistency of the 2 MGD experts, the correlation of the grades by the 2 MGD experts was calculated using the confusion matrix approach.\cite{RN28} With this approach, the mean meiboscore assigned by the MGD experts was defined as the closest integer to the average of 3 gradings by the MGD expert. Simple accuracy was calculated based on the same score prediction. Inter- and intra-rater agreement and disagreement were calculated based on kappa scores.\cite{RN29} \par

\subsection{Deep learning model and training procedure: Segmentation}
While it is common to use pre-processed images in conventional deep neural networks,\cite{RN30, RN31} it is now possible to utilize more features from original raw images than processed images with advanced DL.\cite{RN32} Thus, we used original infrared MG images with investigators' annotated segmentations to train DL. \par

We designed an encoder-decoder based DL model (Fig. \ref{Figure-2.png}) with skip connections and residual connections (ResNet 34) on the encoder part.\cite{RN33, RN34} We have experimented with different encoders (Original U-Net encoder, ResNet 18, ResNet 34, and ResNet 50 encoder) and found that ResNet 34 was the best fit to handle full-size images and provided high accuracy in case of segmentation. The input and output size were single-color channel grayscale images of resolution 1280 \texttimes{} 640 pixels. We used image augmentation methods for image transformation such as `shear' and `horizontal flip up to 15\%'\cite{RN35} without changing the original input image size. In total, we used 1000 images (800 for training and 200 for validation) paired with the corresponding 1000 images of MG segmentations and 1000 images of eyelid segmentation.\par

We initially trained our DL model with the 800 input images (385 upper eyelids, 415 lower eyelids) and 800 MG segmentation annotated by investigators. The same architecture was then trained with the original input images with eyelid segmentations. The DL model was made with the PyTorch framework,\cite{RN36} and then trained on a machine with 1 GPU (GEFORCE RTX 2080 Ti) for 200 epochs with the MGD-1K dataset, which took approximately 24 hours. Training time was not a limitation in our case because, for prediction, the model needs to be trained only once.\par

After our DL model was trained, it could segment out each MG and eyelid from 200 new unseen images (82 upper eyelids, 118 lower eyelids) without any additional help from investigators or MGD experts. First, the MGD experts directly observed the results to check whether there were areas of insufficient or excess markings by DL. Next, MG area and eyelid area were calculated from the MG segmentation and eyelid segmentation. Then MG ratio was calculated from the ratio of MG area and eyelid area.\par

For validation of the DL model based on MG area/ratio, we compared the mean MG area/ratio obtained by investigators and that obtained by the DL using a paired t-test (only 91 left eye images). The correlation between the MG area/ratio by DL model and the MG area/ratio by annotation of the investigators was analyzed (linear regression). We further performed a Bland-Altman analysis. To validate the DL output, we analyzed the correlation between the MG area/ratio and age in the upper and lower eyelids (linear regression), because it is well known that MG dropout increases with age.\cite{RN4} \par

To validate the DL mode, we analyzed the correlation between the DL segmentation-based MG area/ratio and meiboscore determined by the MGD experts. For 189 meibography images, we assigned the arithmetic mean of all 6 rounds of meiboscores. In this way, we obtained a meiboscore with intermediate values (0, 0.17, 0.33, 0.50, \ldots{} 2.83, 3.00). We found the correlation between the DL-based MG area/ratio and the arithmetic mean of all 6 rounds of meiboscores. \par

\subsection{Segmentation accuracy and meiboscore prediction}

\begin{suppfigure}
    \centering
    \includegraphics[width=.70\columnwidth]{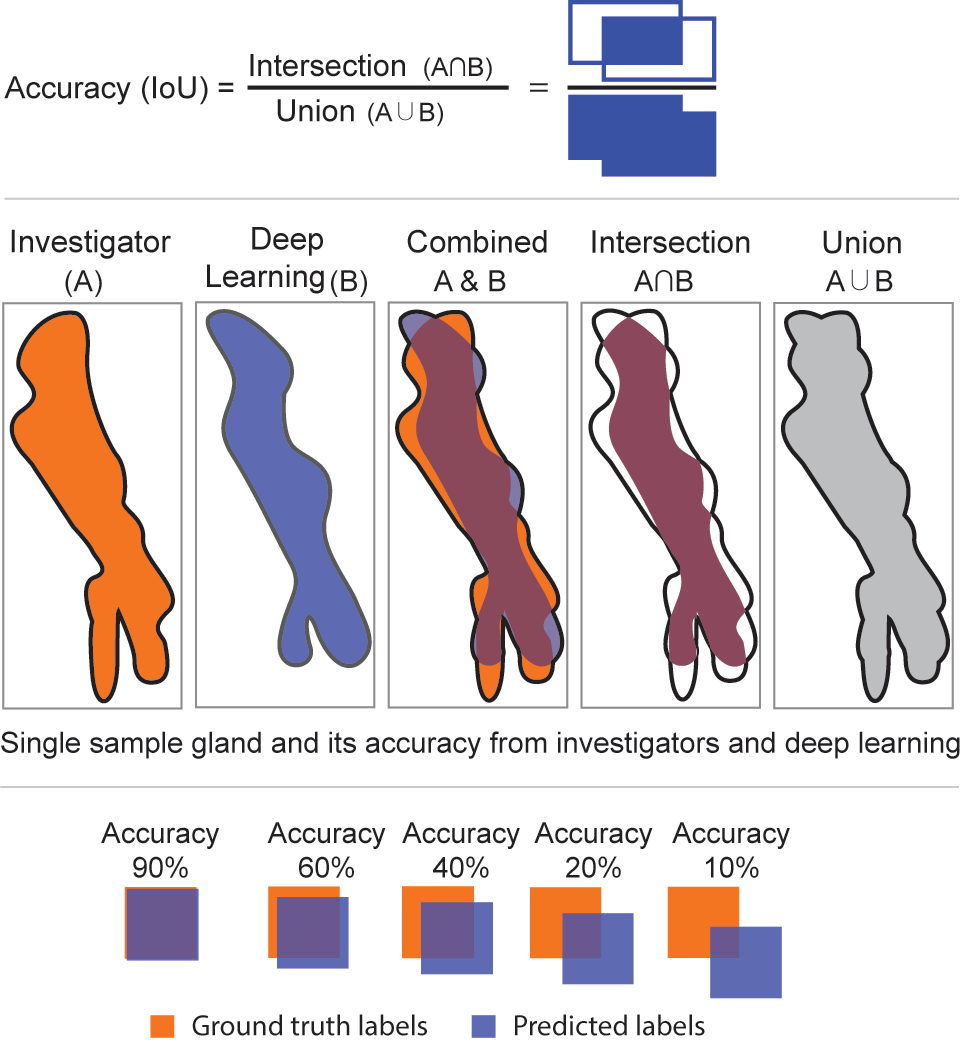}
    \caption{Definition of accuracy (IoU). Accuracy (IoU) is computed by dividing the intersection of 2 images by the union of the 2 images. For example, a picture was labeled by 2 investigators A and B. There is a lot of overlap between A and B. The overlapping area of the 2 images $(A \cap B)$ divided by combining A and B $(A \cup B)$ indicates the accuracy between the 2 segmentation. This accuracy is also known as the intersection over union (IoU), Jaccard index, or Jaccard similarity coefficient. This is the most common measurement system used to measure image segmentation accuracy.}
    \label{Supplement-Figure-1.png}
\end{suppfigure}

We evaluated the segmentation accuracy of the MG and eyelid area at the pixel level based on Intersection over Union (IoU). The IoU measures how much the DL-predicted binary images correspond to the ground truth images by dividing the IoU of these 2 images.\cite{RN37} Similarly, the IoU can be used to measure the match between 2 investigators. The principle of the IoU accuracy matrix in our system is represented in (Supplement Fig. \ref{Supplement-Figure-1.png}. We tried to determine the performance of our DL model compared with that of the investigators. For this, all 200 images of the validation dataset were labeled twice by investigators. We computed the accuracy of the DL by taking the IoU between the `DL-predicted images' and `images annotated by any 2 investigators' (for each 200 validation images, the investigator was randomly selected). Investigator accuracy was determined by calculating the IoU of each marked image pair of the validation set. The mean accuracy (IoU) of the investigators was calculated by taking the mean of all accuracies. \par

The meiboscore prediction part of the model was based on the classification model. Three images were used as input: the original image, segmented MG, segmented eyelid (obtained as output from the segmentation model), and the meiboscore was provided as input. Using this classification architecture, it is also possible to estimate meiboscore directly from original images without those segmentation steps (MG/eyelid segmentation image). However, adding those MG and eyelid segmentation provide more insight to the classification architecture and significantly improves meiboscore prediction. The meiboscore output was given a grade of 0, 1, 2, or 3. To compare the meiboscore prediction accuracy of the DL model, we used the confusion matrix. Mean meiboscore by 2 MGD experts (ground truth) was defined as the closest integer to the mean of the 6 gradings by the MGD experts. Meiboscore prediction accuracy was calculated based on the same prediction score. \par

Besides 1,000 images, another dataset of 600 images from Korea University Ansan Hospital was used to test the meiboscore prediction part of the deep learning model. Another 2 MGD experts (Y.S.E, J.Y.K.) graded meiboscores of 600 images according to Arita's criteria. All 600 images were labeled 3 times for meiboscore by 2 MGD experts separately; thus, 6 times per image. \par

\subsection{Correction of specular reflections by the generative adversarial network model}

\begin{figure}
    \centering
    \includegraphics[width=.99\columnwidth]{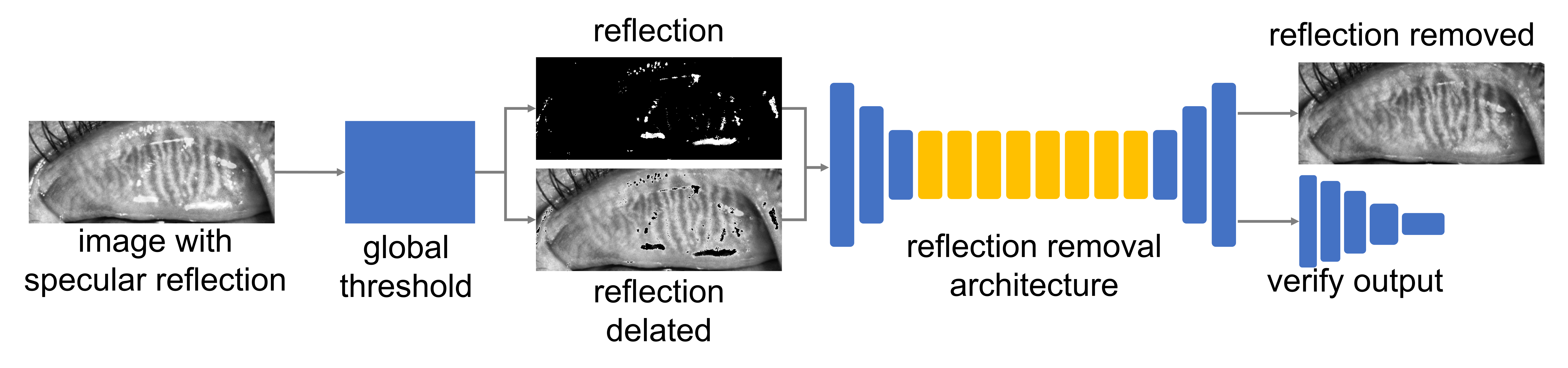}
    \caption{Specular reflection-removal architecture. The original image was first converted to a reflection section and reflection-deleted image based on bright pixel values. The images were then passed into a deep learning model that used a generative adversarial network to restore the reflection portion based on an understanding of the surrounding image area. The network was trained with style loss, perception loss, reconstruction loss, and GAN loss. As an output, the neural network restored the reflection parts with realistic pixel combinations.}
    \label{Figure-3.png}
\end{figure}

Specular reflection is a distracting factor for MGD experts when diagnosing MG from original images. Specular reflection often creates confusion for MG assessment because of the white appearance with saturated pixels. So, we found that besides providing important parameters like MG area/ratio and meiboscore, it is also helpful to remove specular reflection from original images for MGD experts. Different DL-based image restoration model have emerged emerged, \cite{RN38, RN39, RN40} we also applied the DL approach to correct specular reflections from MG images. Specular reflections are usually represented by a high light intensity, i.e., a high pixel value, in the images. Based on the principle of global thresholding, we detected all specular reflections in the MG images. Removing specular reflection is challenging, however, and this problem can be resolved with a generative adversarial network (GAN)-based DL model.\cite{RN41} The GAN architecture comprises a generator and a discriminator component. A variation of the model was implemented to fill the detected specular reflection component of the MG images with possible gland/eyelid portions based on an understanding of surrounding information provided by the whole image.\cite{RN42} This model takes 2 images as input (original image where the reflection portion was filled by a solid color and a specular reflection mask image) and provides 1 output layer \textemdash{}the predicted images without reflection (Fig. \ref{Figure-3.png}). The model was pre-trained with different types of publicly available image datasets to gain the knowledge to restore any missing parts of those images. The generator fills the specular reflection component, and the discriminator verifies the final output repeatedly until the output reflection-free image achieves the best algorithmic performance. This approach remove specular reflection and fill that with understanding of image to make meibography images less distracting to be graded by ophthalmologist. A paired t-test was calculated based on the 109 validation images(only left eye) to represent the difference in the MG area/ratio predicted from a DL algorithm with original images as input vs. specular reflection-corrected images as input. \par

For statistical analysis, GraphPad Prism 8.4.2 (GraphPad Corp., San Diego, CA) was used. A p-value of less than 0.05 was considered significant. \par

\section{Results}
\subsection{Patients and dataset of MGD-1K}

\begin{suppfigure}
    \centering
    \includegraphics[width=.70\columnwidth]{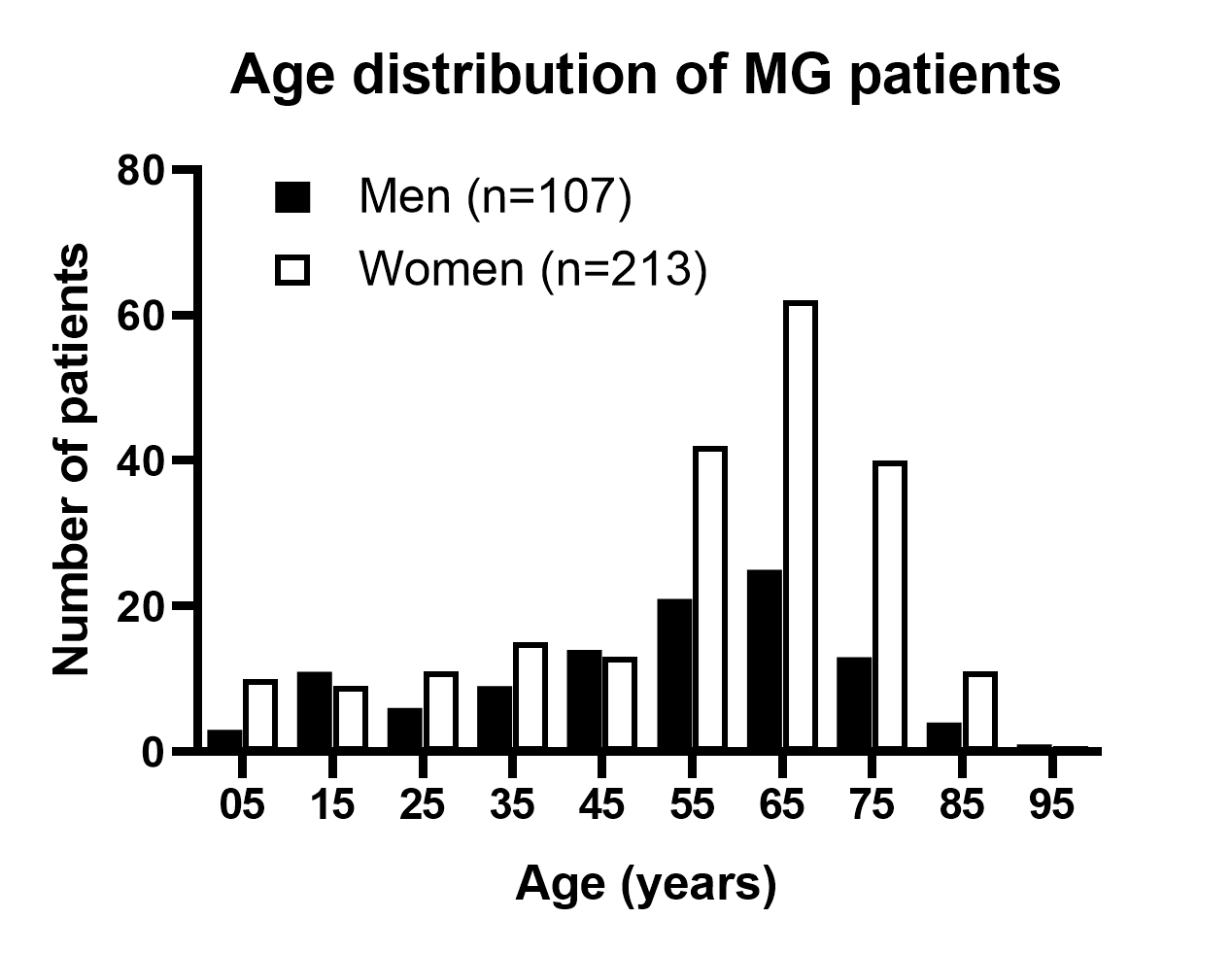}
    \caption{Age distribution of patients with dry eye syndrome included in the study. This graph represents the distribution of the age by sex. The mean age of female patients was 54.14 years (SD 19.64, median 59), and the mean age of male patients was 54.11 years (SD 19.64, median 59).}
    \label{Supplement-Figure-2.png}
\end{suppfigure}

The demographic information of the patients included in the study is presented in Table 1. The mean (SD) age of the 320 patients was 54.6 (20.2) years with 107 men and 213 women (Supplement Fig. \ref{Supplement-Figure-2.png}).\par

\subsection{MG annotation by investigators and meiboscore grading by 2 MGD experts}

\begin{suppfigure}
    \centering
    \includegraphics[width=.99\columnwidth]{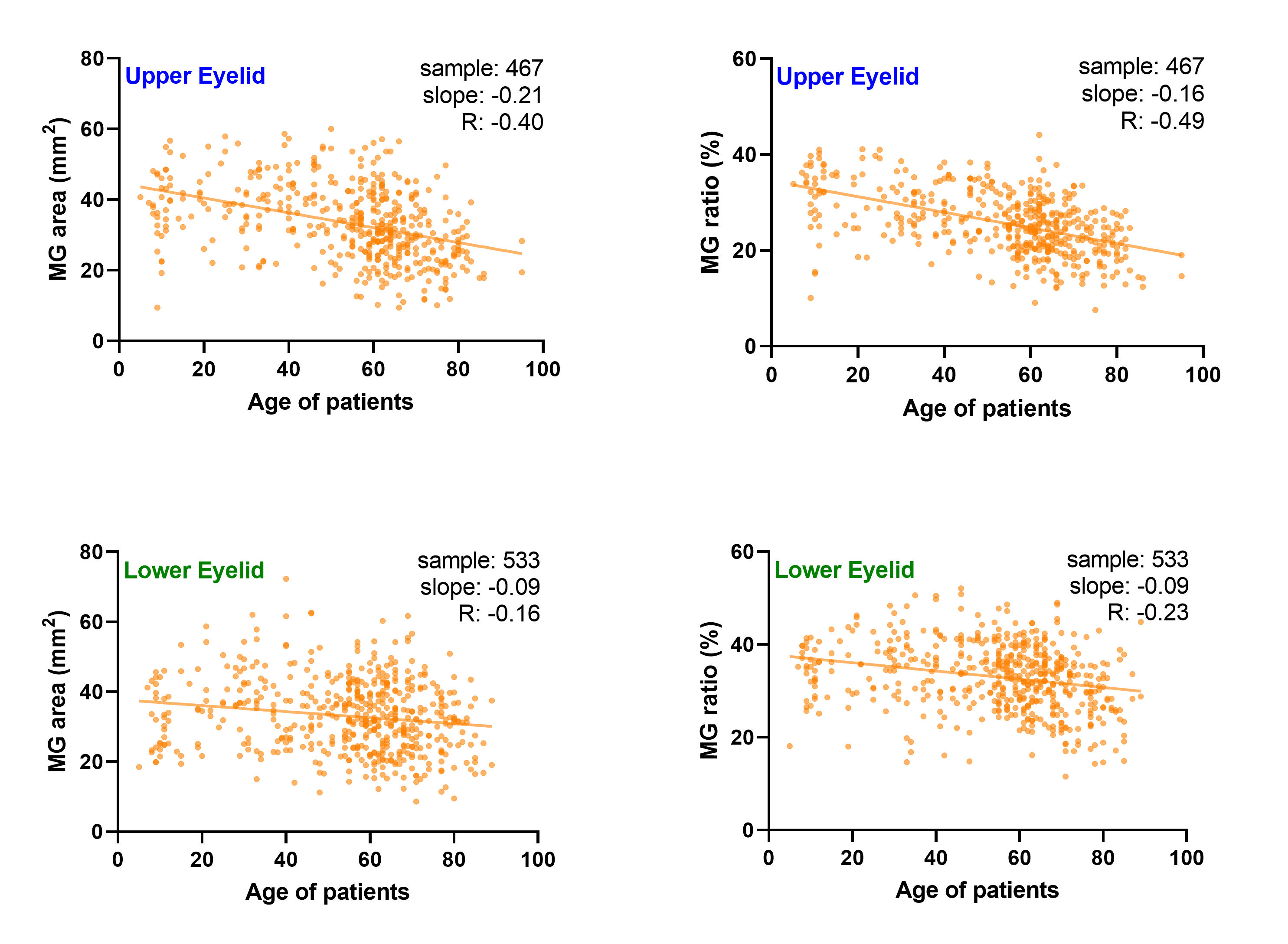}
    \caption{Correlation of the MG area/MG ratio determined by investigators with patient age. The relation is represented for both the upper eyelid and lower eyelid. The correlation indicates that MGD is more common in older people.}
    \label{Supplement-Figure-3.png}
\end{suppfigure}

It took approximately 12\textendash{}25 minutes for the investigators to precisely annotate each MG image with Photoshop. The mean MG area was $33.80 \pm 9.48$ $mm^2$ in the upper eyelid and $32.04 \pm 10.87$ $mm^2$ in the lower eyelid, and the mean MG ratio was $26.23 \pm 5.83\%$ in the upper eyelid and $32.34 \pm 7.45\%$ in the lower eyelid. There was a significant negative correlation between the MG area and age in the upper (p \textless{} 0.0001) and lower eyelids (p = 0.0002), and between the MG ratio and age in the upper (p \textless{} 0.0001) and lower eyelids (p \textless{} 0.0001); (Supplement Fig. \ref{Supplement-Figure-3.png}).\par

\begin{figure}[!ht]
    \centering
    \includegraphics[width=.99\columnwidth]{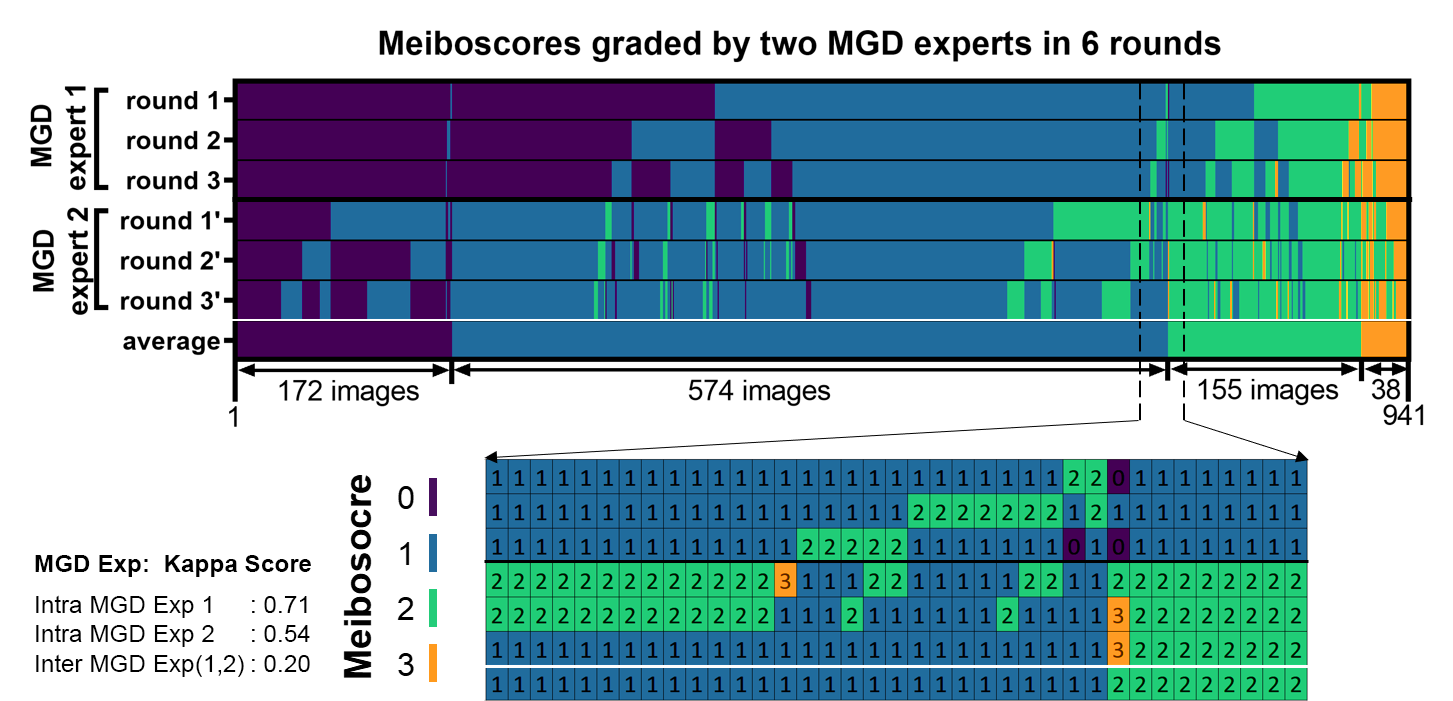}
    \caption{Intra-and intervariability between the meibomian gland dysfunction (MGD) experts. The X-axis represents 941 individual images graded by 2 MGD Experts (ophthalmologists), and the Y-axis represents the meiboscore. Each MGD expert graded the meiboscore 3 times. The first 3 rounds in the Y-axis represent the meiboscore of the first MGD expert (collected with a 1-week inter-scoring interval). Similarly, the next 3 rows represent the meiboscore by the second MGD expert, and in the end, the mean meiboscore is represented by averaging all 6 rounds of meiboscores and rounding up to the next integer. This average meiboscore is used to train and validate the deep learning model and to measure performance. Intergrading accuracy between MGD Expert 1 and 2 are represented by kappa score of 0.20 representing slight agreement. Intra-grading accuracies of MGD Expert 1 and 2 were Kappa scores of 0.71 and 0.54, respectively.}
    \label{Figure-4.png}
\end{figure}

\begin{figure}
    \centering
    \includegraphics[width=.99\columnwidth]{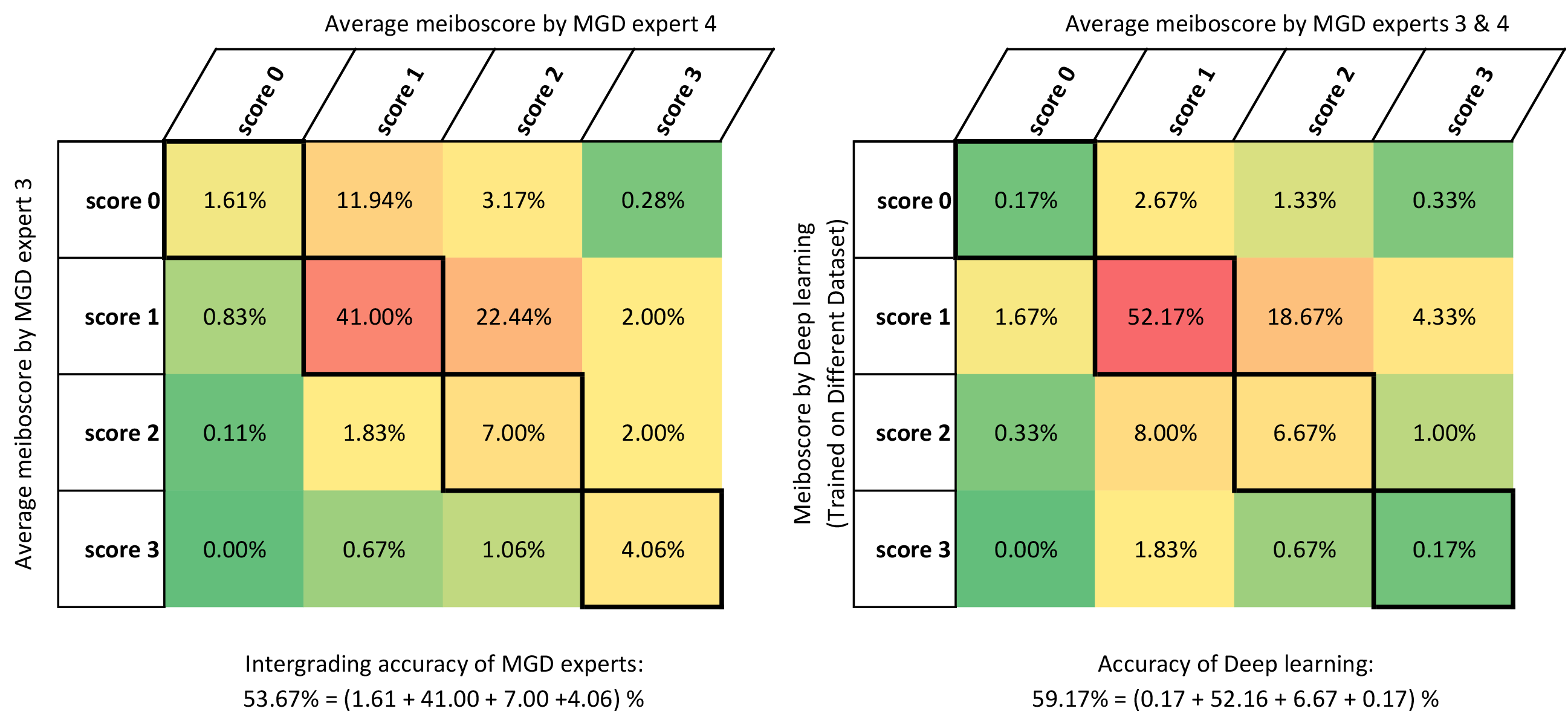}
    \caption{. Meiboscore intergrading accuracy and accuracy of the deep learning. The X-axis and Y-axis represent the category of the meiboscore by the 2 meibomian gland dysfunction (MGD) experts (left side) or deep learning and the MGD expert (right side). The inside values represent the match within those categories. The diagonal line represents the same category matched between the X- and Y-axes.}
    \label{Figure-5.png}
\end{figure}

The scoring was notably inconsistent within and between the 2 MGD experts (Fig. \ref{Figure-4.png}). The color-coded full data representation revealed all grades of 0, 1, 2, and 3 for the same meibography image. Intergrading accuracy between MGD Expert 1 and 2 are represented by kappa score of 0.20 representing slight agreement. Moreover, the intra-grading accuracies of MGD Expert 1 and 2 were Kappa scores of 0.71 and 0.54, respectively. We found a 53.63\% inter-grading similarity between the MGD experts 3 and 4 (Fig. \ref{Figure-5.png}). \par

\subsection{DL model and training: Segmentation}
\begin{figure}
    \centering
    \includegraphics[width=.99\columnwidth]{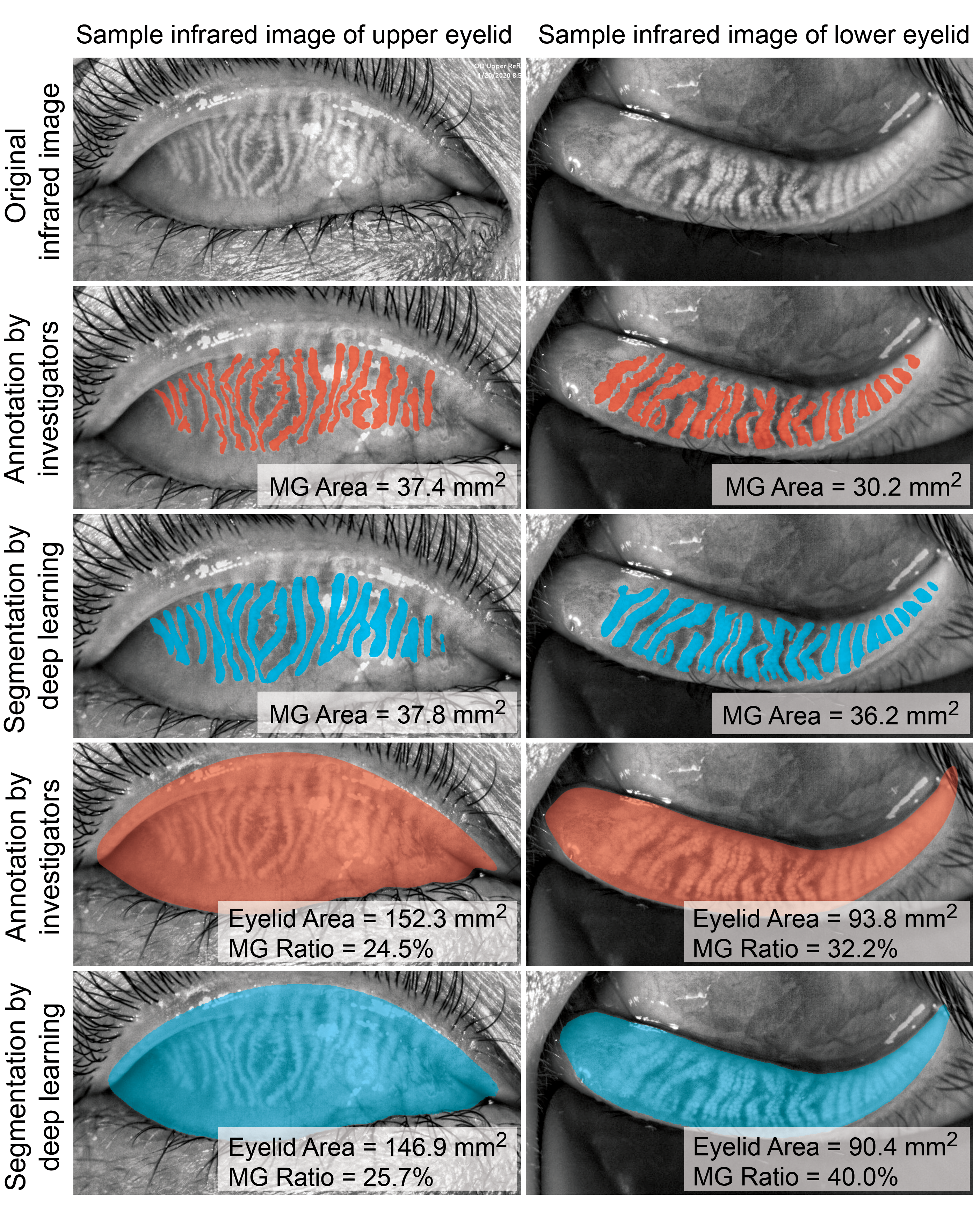}
    \caption{Annotation and prediction. Investigator's segmentation and deep learning output segmentation results are shown for both meibomian glands and eyelids. The area is also shown based on the segmentation. The red transparent color represents the annotation by investigators, and the blue color represents the output segmentation by the deep learning model.}
    \label{Figure-6.png}
\end{figure}

After the DL-based segmentation model was trained with the input images, any grayscale MG infrared image could be transformed into images containing the segmented labels of the MG or eyelid in 0.3s. MGD experts checked the results to determine whether there were areas of insufficient or excess markings, and found no obvious incorrectly marked areas in any of the 1000 images. A representative example of the original infrared image and images overlaid by the investigator's annotation and DL segmented output for the upper and lower eyelid is shown in (Fig. \ref{Figure-6.png}). \par

\begin{figure}
    \centering
    \includegraphics[width=.99\columnwidth]{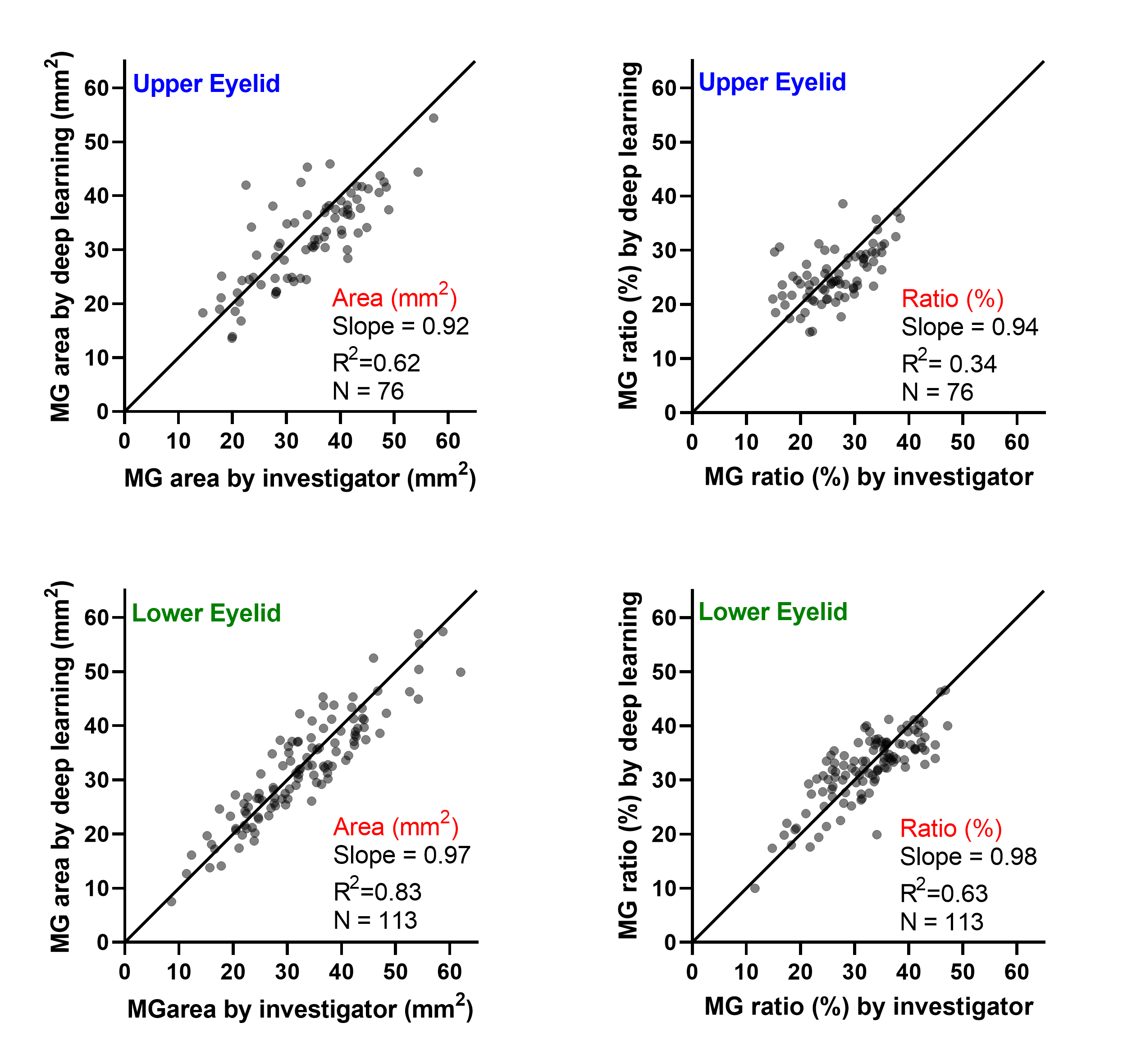}
    \caption{Meibomian gland (MG) area of investigator vs. deep learning model. Correlation of the MG area (mm2) and MG ratio (\%) calculated from the annotations of the MG area by investigators and deep learning.}
    \label{Figure-7.png}
\end{figure}

\begin{figure}
    \centering
    \includegraphics[width=.99\columnwidth]{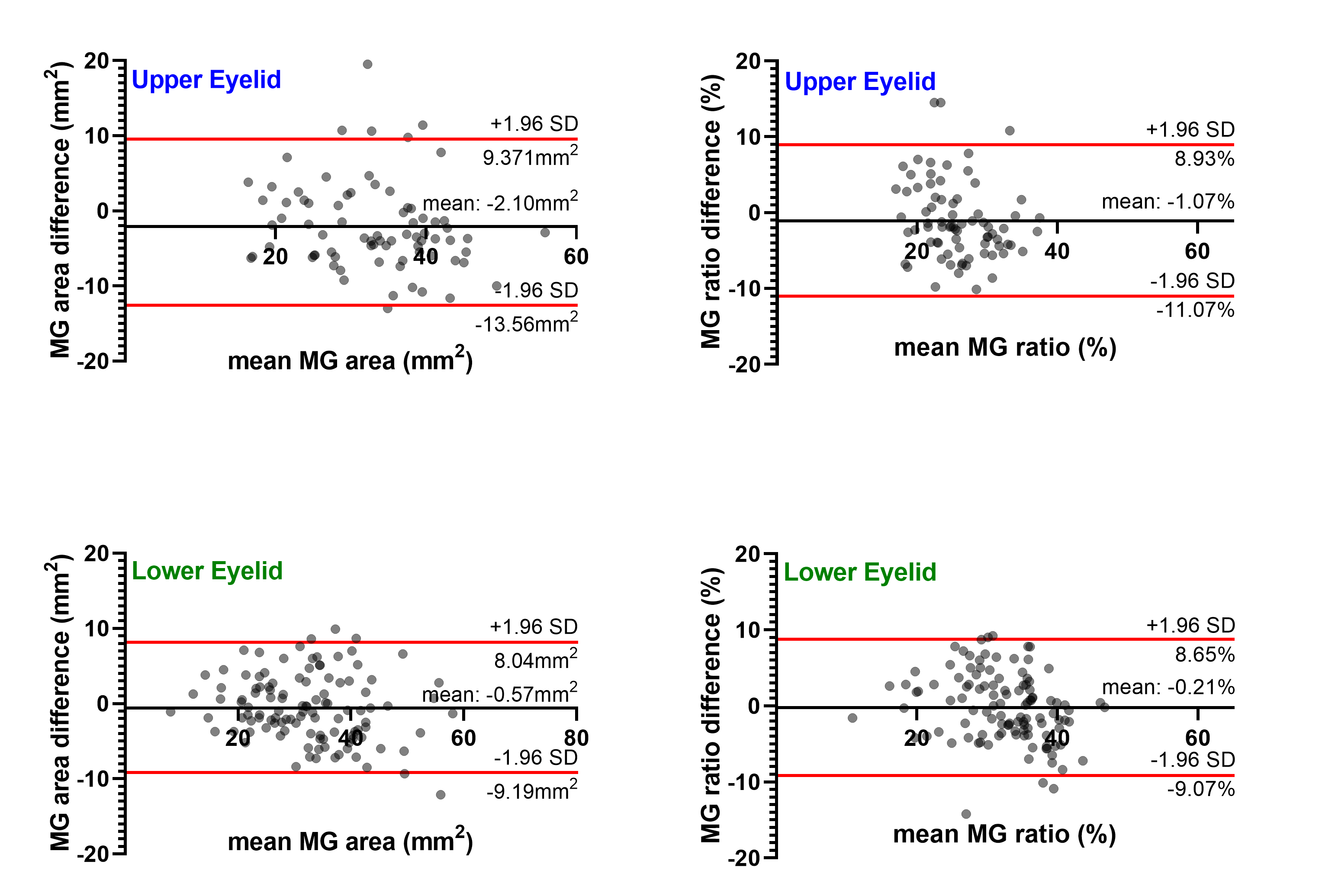}
    \caption{. Bland-Altman plot representing the meibomian gland (MG) area ($mm^2$) and MG ratio (\%) calculated from the annotations of the MG area by investigators and deep learning. The 95\% confidence interval of the differences in the MG area and MG ratio in the upper and lower eyelids of patients was -13.56 to 9.371 and -9.187 to 8.044 in the MG area, respectively, and -11.07 to 8.930 and -9.073 to 8.651 in the MG ratio, respectively. The confidence interval does not reveal any positive or negative bias.}
    \label{Figure-8.png}
\end{figure}

\begin{figure}
    \centering
    \includegraphics[width=.99\columnwidth]{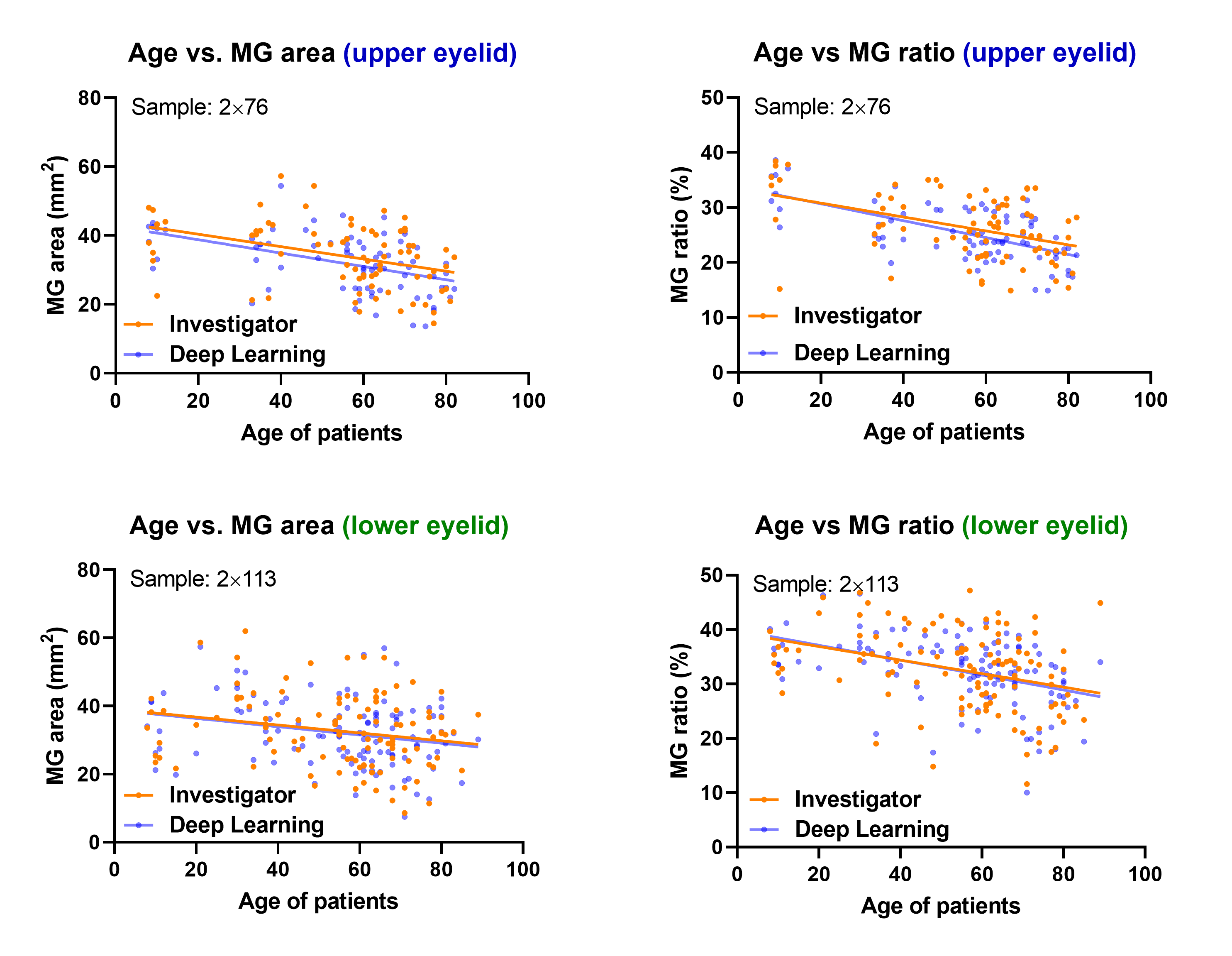}
    \caption{Correlation of the meibomian gland (MG) area/MG ratio with patient age. The relation is represented for both the upper eyelid and lower eyelid by investigators and deep learning. The correlation demonstrates that older people more commonly have MG dysfunction. Usually, the area estimation of deep learning is slightly lower compared to the investigator by a certain threshold.}
    \label{Figure-9.png}
\end{figure}

\begin{figure}
    \centering
    \includegraphics[width=.99\columnwidth]{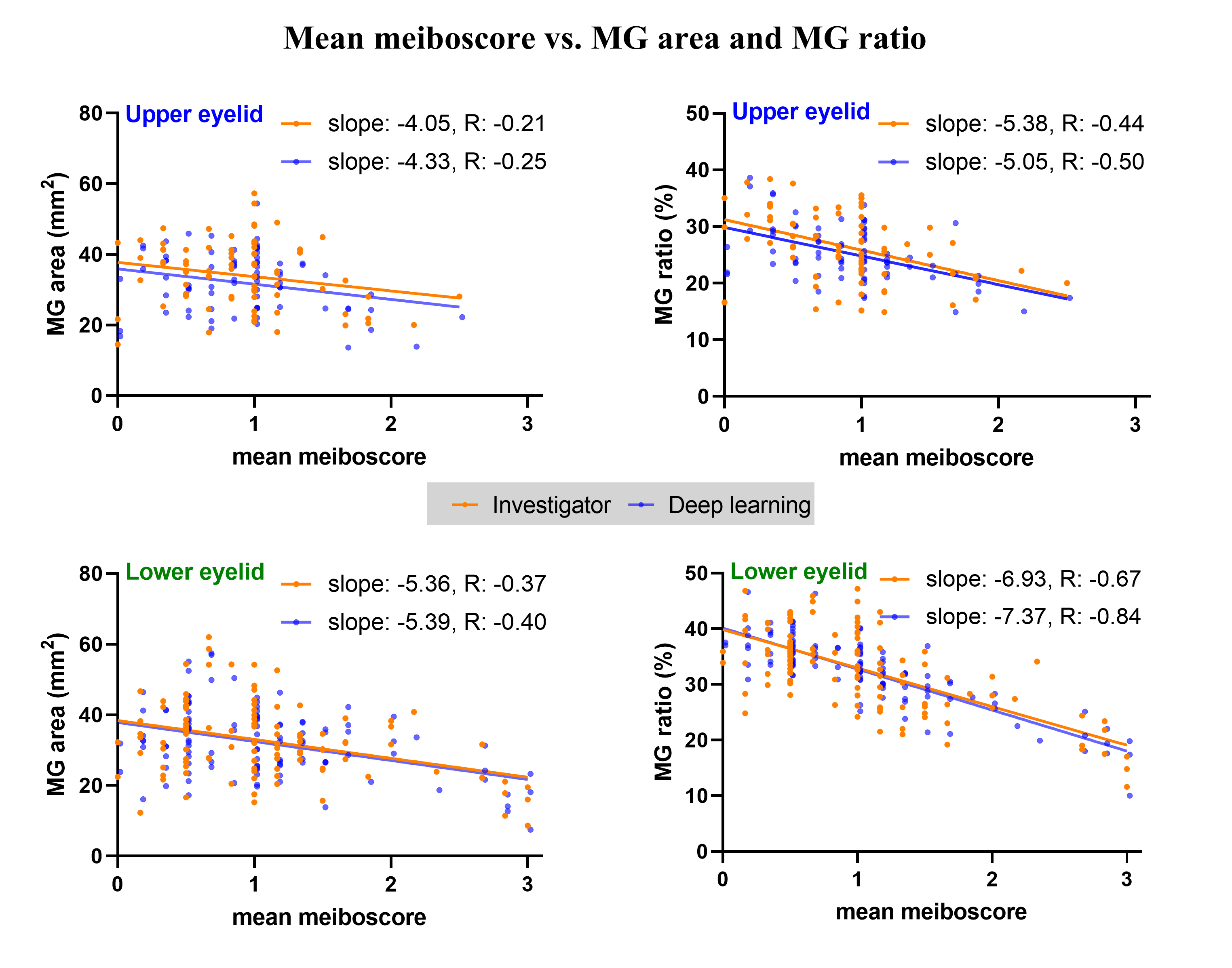}
    \caption{Graph representing the correlation of mean meiboscore from MGD experts and MG area (collected from investigators and deep learning). Investigators and deep learning show a similar correlation in the lower eyelid, but the output area from deep learning is slightly lower for the upper eyelid.}
    \label{Figure-10.png}
\end{figure}

A comparison of MG Area/ratio by investigators and DL-based model is shown with paired t-test based on left eyes in (Table 2). In the upper eyelids,  a paired t-test showed no significant difference between the annotated results obtained by the investigators and the DL results (Table 2). In the upper-lower eyelid, the DL-based MG area/MG ratio was slightly underestimated overestimated compared with the investigator annotated MG area/MG ratio (p = 0.048).  The MG area and MG ratio of both eyelids obtained by DL and the investigators had a good positive correlation (p < 0.0001, Fig. \ref{Figure-7.png}). A Bland-Altman plot representing the MG area (mm2) and MG ratio (\%) calculated from the MG area by investigators and by DL is shown in (Fig. \ref{Figure-8.png}). The plot shows that the 95\% confidence interval of the differences in the MG area and MG ratio in the upper and lower eyelids represented no positive or negative bias. Regarding the effect of age on the MG area and MG ratio, there was a significant negative correlation of the DL-based and investigator-based MG areas and MG ratios in the upper and lower eyelids with respect to age (Fig. \ref{Figure-9.png}). As the arithmetic mean of all 6 rounds of meiboscores increased, the MG area and MG ratio measured by investigators and DL segmentation in the upper and lower eyelids decreased (Fig. \ref{Figure-10.png}). In all 4 cases, the investigator and DL showed similar tendencies with similar linear regression lines. \par

\subsection{Segmentation accuracy and meiboscore prediction}
\begin{figure}
    \centering
    \includegraphics[width=.99\columnwidth]{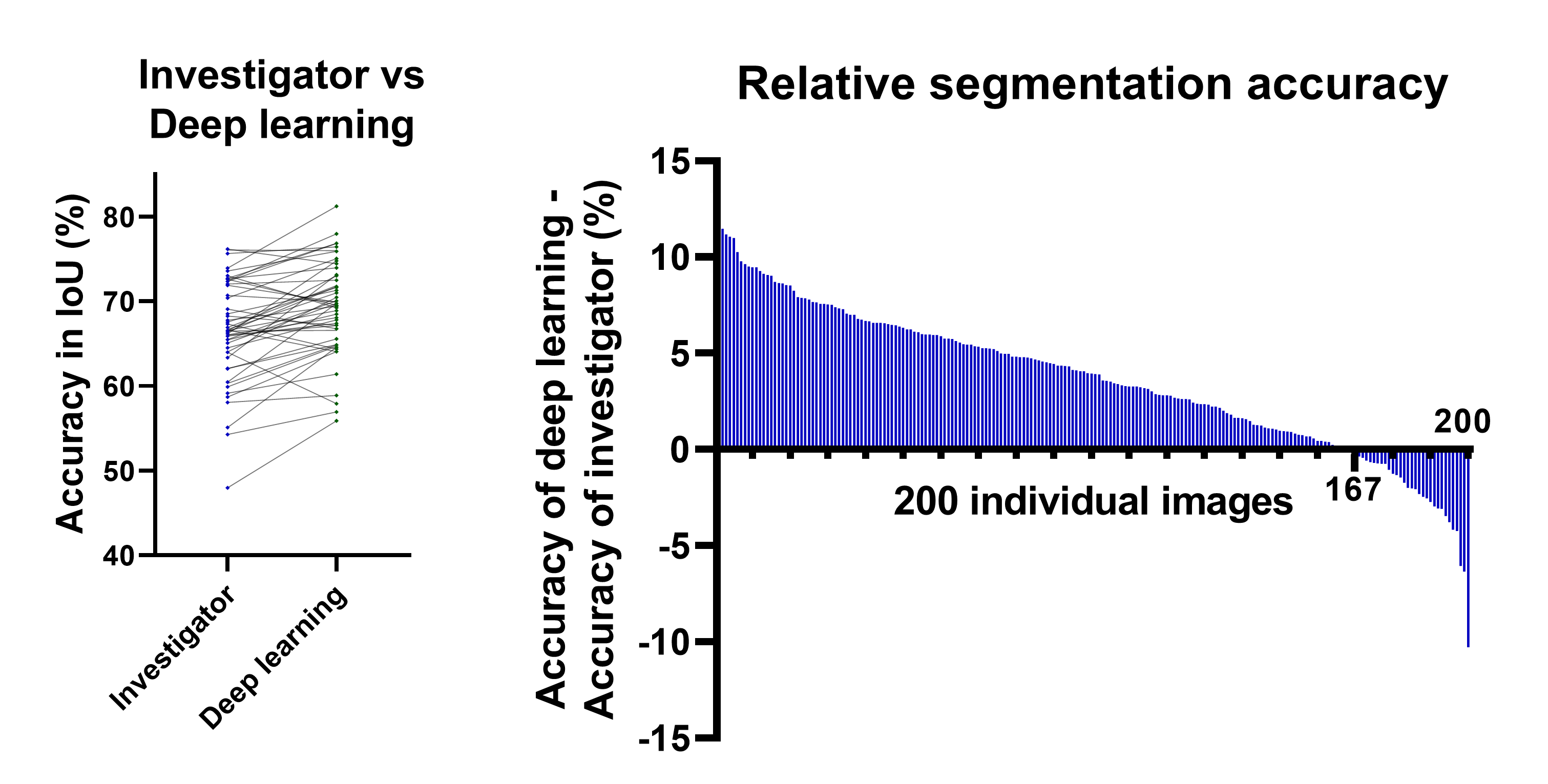}
    \caption{Segmentation accuracy of investigators and deep learning model. The accuracy (IoU) distribution of 50 MG images from 200 images of the validation set is displayed on the left side. In most cases, the deep learning model showed better accuracy segmenting the MG. The relative segmentation accuracy of the deep learning model was determined by representing the difference of Accuracy of Model (IoU) and Accuracy of investigators (IoU) for all 200 validation images (represented on the right side). All 200 validation images were labeled twice by investigator comparison. The ophthalmologists had 64.03\% agreement between those 2 segmented sets (SD 6.453\%). When the deep learning model (trained on different images) predicted segmentation images of those 200 unseen images, there was a 67.63\% match between those predicted images and images marked by the ophthalmologists (SD: 6.635\%). In 167 of 200 images, the machine learning algorithm had higher accuracy segmenting those glands compared with the investigators.}
    \label{Figure-11.png}
\end{figure}

The accuracy of the investigators was computed by taking the IoU between the 2 investigators. The accuracy of the investigators was 64.03\% (Fig. \ref{Figure-11.png}). With the 200 validation images, the DL model's accuracy (mean IoU score) was 67.63\% for the MG output. \par

Regarding meiboscore prediction, we found a 73.01\% agreement between the DL-predicted meiboscore and the mean of 6 gradings by MGD experts 1 and 2  (Supplement Fig. \ref{Supplement-Figure-4.png}). The intergrading agreement between MGD expert 1 and 2 was 53.44\%. However, when the pre-trained model was used to predict meiboscore from an independent dataset of 600 images with meiboscore from MGD expert 3,4, the agreement was not as high as the original dataset (Fig. \ref{Figure-5.png}). The intergrading agreement between MGD expert 3 and 4 was 53.67\%. \cite{RN43} (Supplement Fig. \ref{Supplement-Figure-5.png}) shows deviation of MGD experts and deep learning prediction of Meiboscore. \par

\begin{suppfigure}
    \centering
    \includegraphics[width=.99\columnwidth]{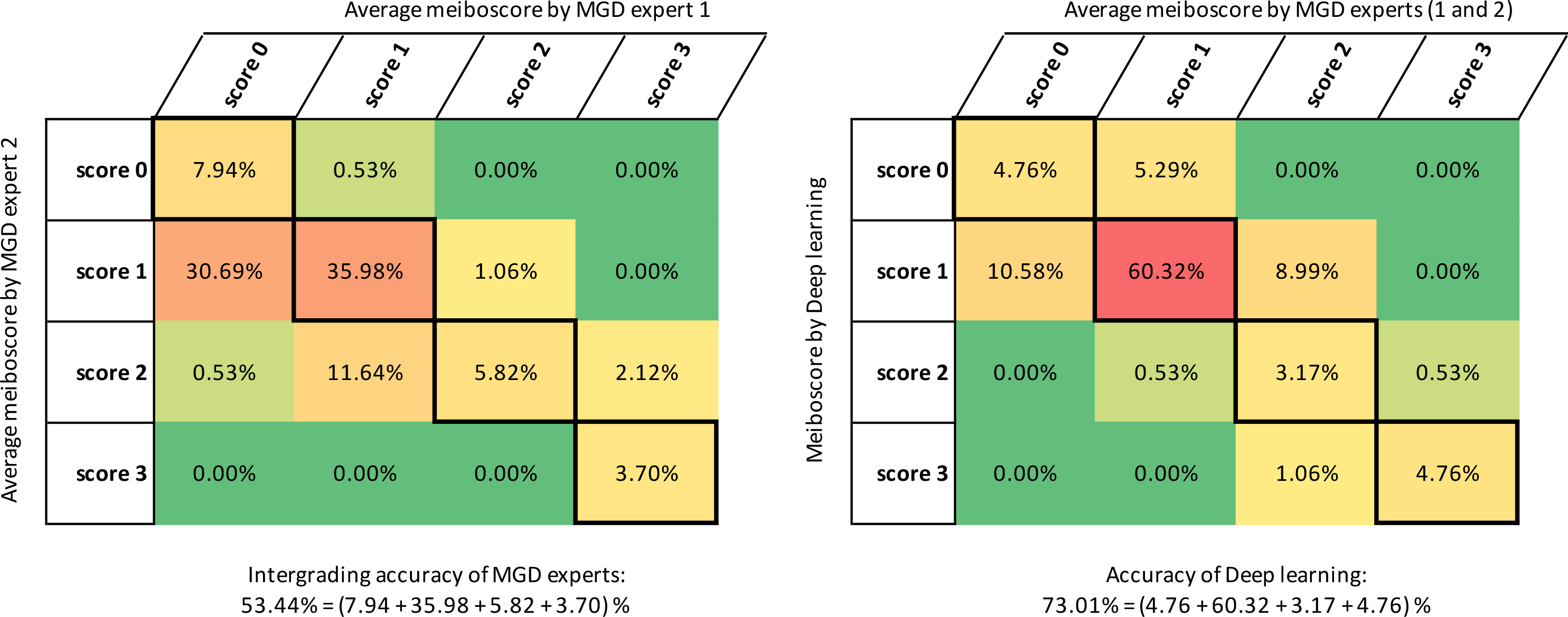}
    \caption{Meiboscore intergrading accuracy and accuracy of the deep learning. The X-axis and Y-axis represent the category of the meiboscore by the meibomian gland dysfunction (MGD) experts 1 and 2 (left side) or deep learning and the MGD expert (right side). The inside values represent the match within those categories. The diagonal line represents the same category matched between the X- and Y-axes.}
    \label{Supplement-Figure-4.png}
\end{suppfigure}

\begin{suppfigure}
    \centering
    \includegraphics[width=.99\columnwidth]{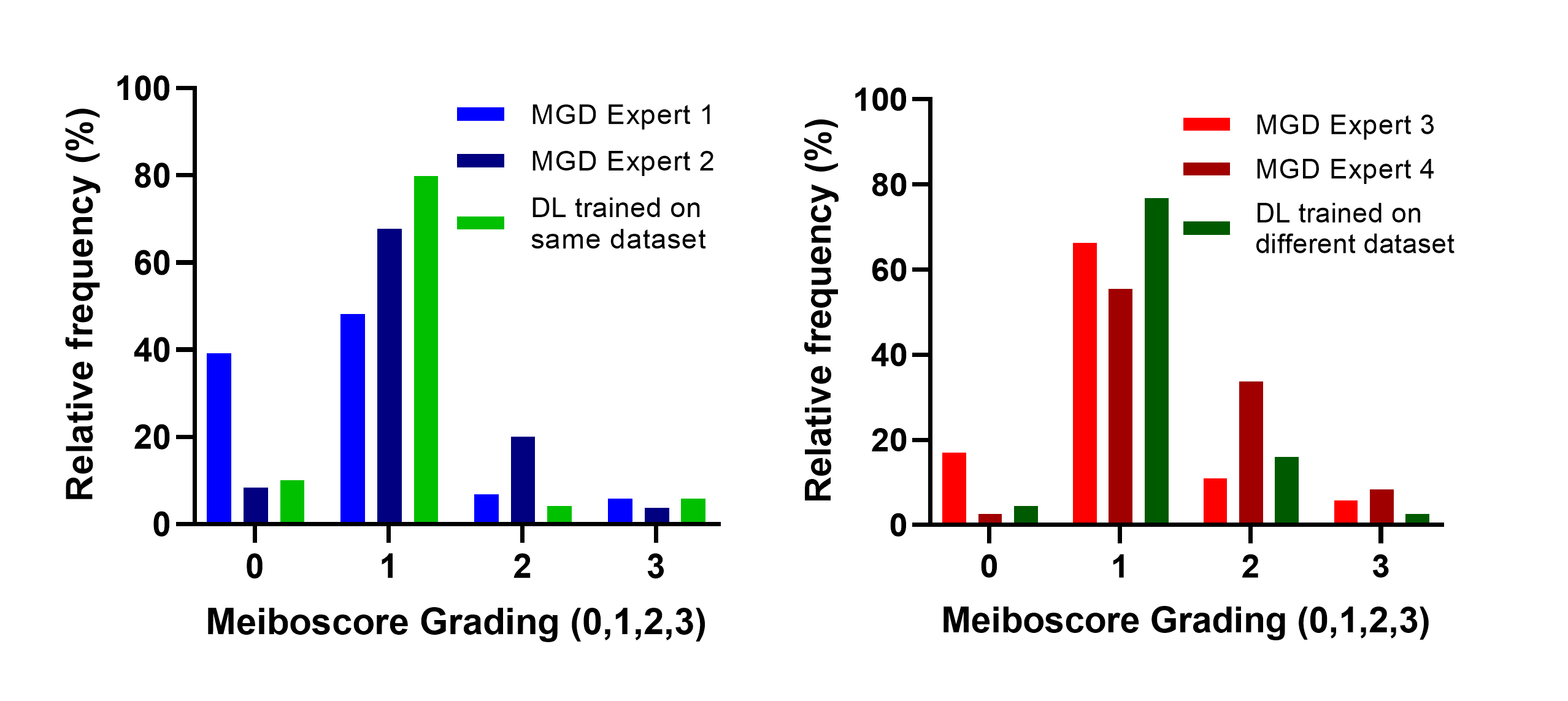}
    \caption{Deviation of meiboscore prediction from deep learning compared to MGD experts. The Y-Axis represent percentage of 4 meiboscores by each MGD experts or deep learning. The deep learning model was only trained on 800 images with meiboscore from MGD Expert No 1 and 2 and tested on both validation images(from MGD Expert 1,2) and on 600 images from an independent center(from MGD Expert 3,4) without retraining again. We see that the meiboscores from different MGD experts were not consistent and high variability can be seen with meiboscore 0, 2 and 3. Deep learning seems to predict more number of meiboscore 1 compared to 0,2,3.}
    \label{Supplement-Figure-5.png}
\end{suppfigure}

\subsection{Correction of specular reflection region based on DL}
\begin{figure}
    \centering
    \includegraphics[width=.99\columnwidth]{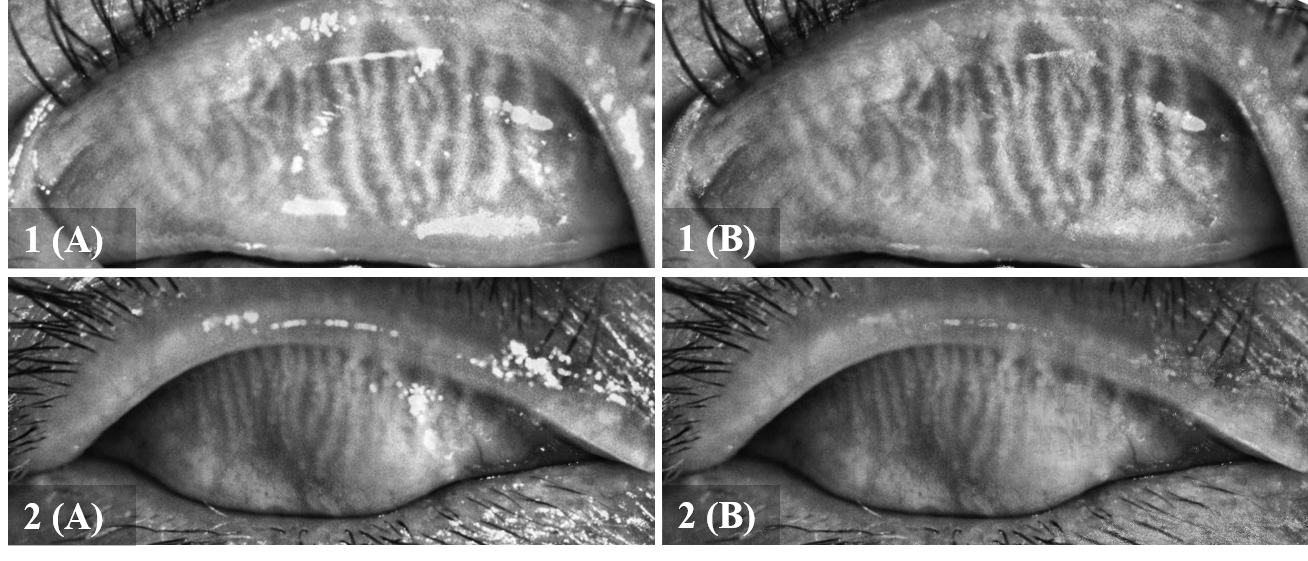}
    \caption{Reflection-removed images. 1(A) and 2(A) represent magnified views of original images and 1(B) and 2(B) represent their corresponding reflection-corrected images. The deep learning model was able to remove specular reflection in most of the cases. As reflection is usually represented by the bright color region, it can be detected by a simple global threshold. Then the deep learning model was applied to remove reflections.}
    \label{Figure-12.png}
\end{figure}

The DL-based approach was able to remove reflection regions from the original MG images and fill that based on the understanding of the whole image. The output seems natural (Fig. \ref{Figure-12.png}). Also, we didn't lose any image details when using this specular reflection process. When we tested original images and reflection-removed images to find the MG area/ratio, there was no significant difference in the MG area (p = 0.689) and MG ratio (p = 0. 255) between the segmented output from the original and reflection-free MG images (Table 3). This justifies that ML-based reflection removal provides not only natural-looking images but also reliable prediction results without deviation. \par

\section{Discussion}

In present study, we demonstrated that DL can automatically detect all individual MGs and quantify the MG area and MG area ratio. Other studies have evaluated methods to quantify MG dropout in meibography.\cite{RN14, RN15, RN16, RN17, RN18, RN20, RN21, RN22} Among various types of commercial equipment, Idra (SBM Sistemi, Orbassano, Italy) shows meibography and the quantified dropout ratio. Recently, Vigo et al. \cite{RN44}

evaluated the diagnostic performance of Idra, including measurement of MG loss, but they did not describe how they quantified and validated MG loss. We could find no other literature referencing the Idra system. \par

\begin{figure}
    \centering
    \includegraphics[width=.99\columnwidth]{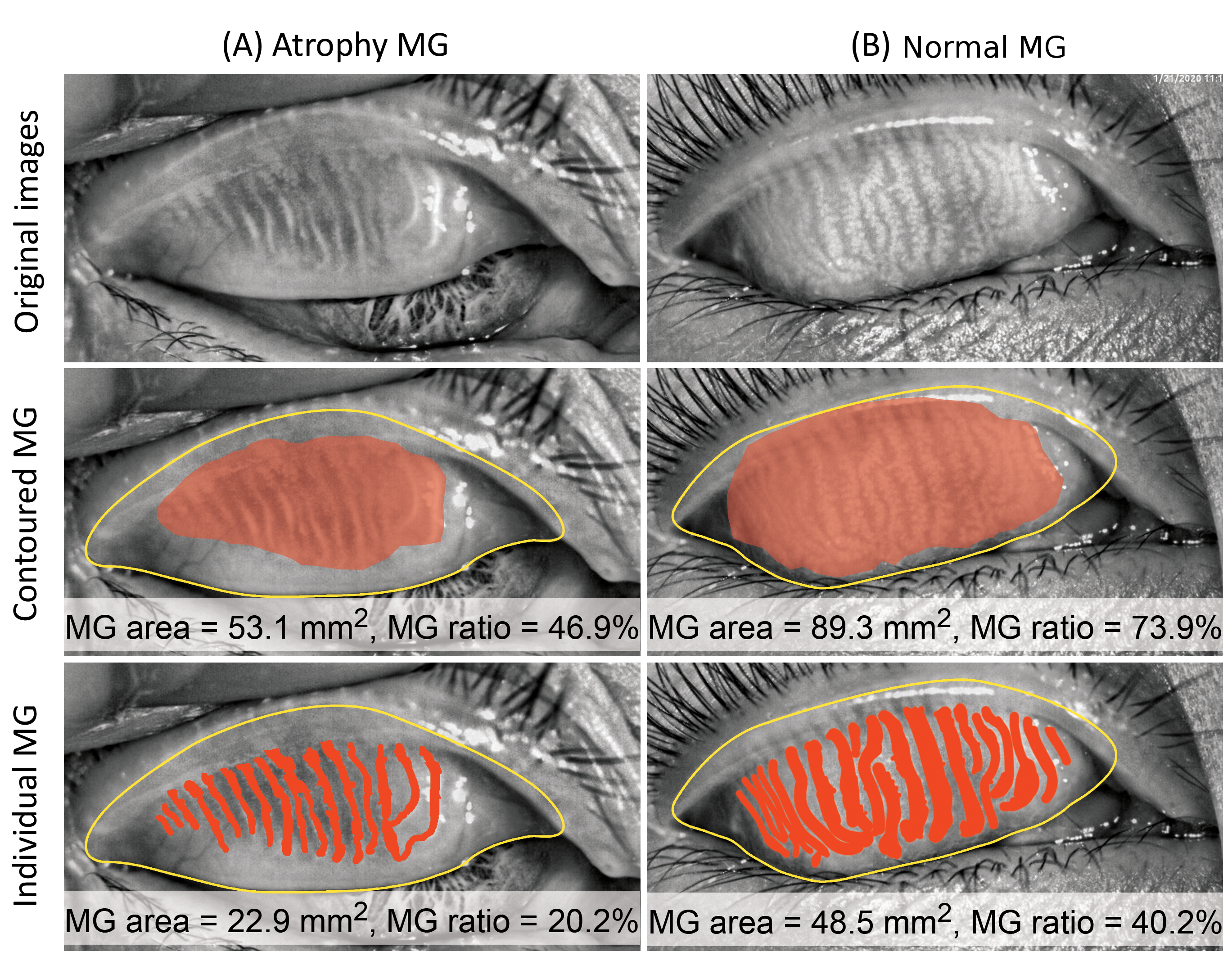}
    \caption{. Comparison of the meibomian gland (MG) area and the MG ratio between contoured MG and individual MG analyses in atrophic MG (A) and normal MG (B). The MG area of atrophic MGs and normal MGs calculated by contoured MG analysis was 53.1 mm2 and 89.3 mm2, respectively. The MG area of atrophic MGs and normal MGs calculated by individual MG analysis was 22.9 mm2 and 48.5 mm2, respectively. The actual MG area calculated by individual MG analysis was much smaller than the area calculated by contour MG analysis. Furthermore, the MG area of normal MGs (89.3 mm2) was only 1.68-fold larger than that of atrophic MGs (53.1 mm2) by contour MG analysis. The MG area of normal MGs (48.5 mm2) was 2.12-fold larger than that of atrophic MGs (22.9 mm2) by individual MG analysis. Therefore, the individual MG analysis is more accurate and better represents MG atrophy than contoured MG analysis.}
    \label{Figure-13.png}
\end{figure}

This study demonstrated the capability of DL to automatically detect all individual MGs and calculate the MG area and MG ratio. When quantifying MG dropout, it may be more meaningful to calculate the whole area of individual MGs rather than to calculate on the basis of a drawn contour of the MG area without dropout, as done in previous papers.\cite{RN14, RN15, RN16, RN17} For example, in (Fig. \ref{Figure-13.png}), Comparison of the meibomian gland (MG) area and the MG ratio between contoured MG and individual MG analyses in atrophic MG (A) and normal MG (B). The MG area of atrophic MGs and normal MGs calculated by contoured MG analysis was 53.1 mm2 and 89.3 mm2, respectively. The MG area of atrophic MGs and normal MGs calculated by individual MG analysis was 22.9 mm2 and 48.5 mm2, respectively. The actual MG area calculated by individual MG analysis was much smaller than the area calculated by contour MG analysis. Furthermore, the MG area of normal MGs (89.3 mm2) was only 1.68-fold larger than that of atrophic MGs (53.1 mm2) by contour MG analysis. The MG area of normal MGs (48.5 mm2) was 2.12-fold larger than that of atrophic MGs (22.9 mm2) by individual MG analysis. Therefore, the individual MG analysis is more accurate and better represents MG atrophy than contoured MG analysis. Similarly, individual MG images may provide a more accurate estimate of the MG ratio. \par

We measured both the MG area and MG ratio in this study. When calculating the MG ratio, the MG area should be divided by the area where the MG should normally be (i.e., eyelid area). In the upper eyelid, the fold line of the everted eyelid has a definite tarsal plate, so the denominator of the ratio is definite. In the lower eyelid, however, it is difficult to determine exactly where the MG should normally be found.\cite{RN26} Therefore, the MG ratio is somewhat inaccurate for the lower eyelid. So, especially in the lower eyelid, we should measure not only the MG ratio but also the MG area itself. \par

For each meibography image, the DL generated a meiboscore prediction based on the 6 meiboscores by the 2 MGD experts. In fact, because the DL-segmented MG ratio was calculated, the meiboscore could have been predicted according to Arita's criteria\cite{RN4} with the MG ratio result.  When grading with the DL MG ratio, however, the proportion of those with a meiboscore of 0 (no dropout) is almost 0\% by definition. Due to this disadvantage, we used DL-produced meiboscores in this study.Until now, in the semi-quantitative grading systems, a grade of 0 is defined as no gland dropout.\cite{RN4, RN12, RN13} By definition, the proportion of those eyelids with a meiboscore of 0, or no dropout, may not exist in a quantitative grading approach. Therefore, it might be better to designate the intervals equally with grade 1: \textless{}25\%, grade 2: \textgreater{}25\% \& \textless{}50\%, grade 3: \textgreater{}50\% \& \textless{}75\%, and grade 4: \textgreater{}75\%. \par

Regarding the meiboscore grading by experts, the consistency was evaluated by 4 MGD experts in 3 rounds for the same 941 meibography images (Fig. \ref{Figure-4.png}) as well as from another different test center. Comparison of the intraobserver grades revealed an inter-grading aggreement of only 53.67\% (Fig. \ref{Figure-5.png}). Previous studies described similar inconsistencies and weak intergrading correlation in meibography.\cite{RN9, RN16, RN45} On the other hand, the DL-based meiboscore prediction achieved an accuracy of 73.01\% on images annotated by MGD Expert 1,2 and 59.17\% on images annoted by MGD Experts of different test center. \par

In this study, the DL performance was validated by various methods. First, MGD experts directly observed the results to check whether there were areas of insufficient or excess markings. Second, the mean DL-based MG area and MG ratio were compared with those measured by investigators. The correlation between the MG area/MG ratio by the DL model and the MG area/MG ratio by annotation of the investigators was analyzed (linear regression) (Fig. \ref{Figure-7.png}). Also, Bland-Altman analysis was performed (Fig. \ref{Figure-8.png}). Third, the correlation between the MG area/MG ratio and age was analyzed between the investigator and DL results (Fig. \ref{Figure-9.png}). Fourth, the correlation between the MG area/MG ratio with the meiboscores determined by the MGD experts and DL segmentation was compared (Fig. \ref{Figure-10.png}). The findings of each of these methods indicated a good DL performance. \par

Furthermore, for the first time, we applied DL to correct specular reflection from the meibography images. In previous automated detection systems,  specular reflections were mistaken as an MG.\cite{RN9} Instead of excluding all reflections, our DL model could restore reflection regions without noticeable artifacts (Fig. \ref{Figure-12.png}), resulting in specular reflection-free images that retained all gland information for easy and distraction-free images for clinicians. We found no significant differences in the MG area/MG ratio due to specular reflection removal (Table 3), implying minimal data loss. \par

Compared with other studies, the present study has the following advantages. First, we used 1600 images. This was possible only because DL segmentation was completely automated, and not manual. Second, unlike previous studies, the analysis was conducted not only on the upper eyelid but also on the lower eyelid. Third, our designed DL model can produce a segmentation mask in 0.3 second directly from original images without any further input instruction and works for both the upper and lower eyelids. Fourth, because images from Lipiview, a widely used apparatus, are used, they can be commercialized as is if our newly developed software is installed on the apparatus. It is also good for use by other researchers with the same kind of data. Fifth, validation was performed for DL-segmented MG area/MG ratio in several ways. Sixth, we have evaluated the model based on different dataset of 600 labeled images. Seventh, we also implemented a correction of the specular reflection in the MG images using a GAN-based DL model. \par

In conclusion, this study describes an automated method for quantifying individual MGs to identify the MG area, eyelid area, MG ratio, and classify the meiboscore in non-contact infrared images of upper and lower eyelids using a specialized method comprising multiple DL blocks. This fully automated DL approach provides better accuracy in terms of segmentation and prediction of the meiboscore. Further, we created the MGD-1K dataset, which is fully public (https://mgd1k.github.io), as the first image-based dataset on meibomian glands with precise eyelid segmentations, gland segmentation, and meiboscores for open research on meibomian gland dysfunction. \par

\section*{Financial disclosures}
The authors have no financial interest or relationship to disclose. 

\section*{Declaration of competing interest}
The authors declare no conflicts of interest.

\section*{Acknowledgments}
The work was supported by the GIST Research Institute (GRI) research collaboration grant funded by GIST in 2022; The Brain Research Program through the N.R.F. funded by the Ministry of Science, I.C.T. \& Future Planning (NRF-2017M3C7A1044964), and the Korea Medical Device Development Fund grant funded by the Korea government, the Ministry of Science and ICT, the Ministry of Trade, Industry and Energy, the Ministry of Health \& Welfare, the Ministry of Food and Drug Safety (Project Number: 1711138096, KMDF\_PR\_220200901\_0076) to E.C.; the 2022 Joint Research Project of Institutes of Science and Technology to E.C.; a grant from the National Research Foundation of Korea (N.R.F.) funded by the Korean government (MEST) (NRF-2019R1A2C2086003 to E.C.; the Korea Health Technology R\&D Project through the Korea Health Industry Development Institute (KHIDI) funded by the Ministry of Health \& Welfare Republic of Korea (grant number: HI17C0659, HI17C0659) (H.S.H.)$;$ and the Basic Science Research Program through the National Research Foundation of Korea (NRF), funded by the Ministry of Education, Republic of Korea (No. 2020R1A2C1005009) (H.S.H.).

\appendix

 \bibliographystyle{elsarticle-num} 
 \bibliography{cas-refs}





\end{document}